\documentclass[aps, prb, twocolumn, showpacs, superscriptaddress, floatfix]{revtex4-1}
\usepackage[utf8]{inputenc}
\usepackage{amsmath, amssymb}
\usepackage{graphicx}
\usepackage{hyperref}
\usepackage{color}

\usepackage[makeroom]{cancel}
\usepackage{stmaryrd}
\def\be{\begin{equation}}
\def\ee{\end{equation}}

\newcommand{\nd}[1]{ \mathrm{{\textbf{#1}}} }
\newcommand{\hnd}[1]{ \hat{\mathrm{{\textbf{#1}}}} }

\newcommand*{\lists}[2]{\left\llbracket \begin{matrix} #1 \\ #2 \end{matrix} \right\rrbracket}
\providecommand{\e}[1]{\ensuremath{{\times} 10^{#1}}}

\definecolor{orange}{rgb}{1,0.5,0}

\begin{document}

\title{A Quantum Monte Carlo algorithm for out-of-equilibrium Green's functions at long times}
\author{Corentin Bertrand}
\affiliation{Univ. Grenoble Alpes, CEA, INAC-PHELIQS, GT F-38000 Grenoble, France}
\author{Olivier Parcollet}
\affiliation{Center for Computational Quantum Physics, Flatiron Institute, 162 5th Avenue, New York, NY 10010, USA}
\affiliation{Institut de Physique Th\'eorique (IPhT), CEA, CNRS, UMR 3681, 91191 Gif-sur-Yvette, France}
\author{Antoine Maillard}
\affiliation{Laboratoire de Physique de l’Ecole Normale Sup\'erieure, UMR 8550, 24 Rue Lhomond, 75005 Paris, France}
\author{Xavier Waintal}
\affiliation{Univ. Grenoble Alpes, CEA, INAC-PHELIQS, GT F-38000 Grenoble, France}

\begin{abstract}
We present a quantum Monte-Carlo algorithm for computing the perturbative
expansion in power of the coupling constant $U$ of the out-of-equilibrium Green's functions of interacting
Hamiltonians of fermions.
The algorithm extends the one presented in {\it Phys. Rev. B 91 245154 (2015)}, and inherits its main property:
it can reach the infinite time (steady state) limit
since the computational cost to compute order $U^n$ is uniform versus time; the computing time increases as $2^n$.
The algorithm is based on the Schwinger-Keldysh formalism and can be used for both
equilibrium and out-of-equilibrium calculations. It is stable at both small and
long real times including in the stationary regime, because of its automatic
cancellation of the disconnected Feynman diagrams.  We apply this technique to the
Anderson quantum impurity model in the quantum dot geometry to obtain the
Green's function and self-energy expansion up to order $U^{10}$ at very low temperature. 
We benchmark our results at weak and intermediate coupling with high precision Numerical
Renormalization Group (NRG) computations as well as analytical results.
\end{abstract}

\maketitle

%%%%%%%%%%%%%%%%%%%%%%%%%%%%%%%%%%%%%%%%%%%%%%%%%%%%%%%%%%%%%%%%%%%%%%%%%%%%%%%
\section{Introduction}

The study of the out-of-equilibrium regime of strongly correlated many-body quantum problems is a subject of growing interest in theoretical condensed matter physics,
in particular due to a rapid progress in experiments with \textit{e.g.} the ability to control
light-matter interaction on ultra-fast time scale\cite{Foerst2011}, light-induced
superconductivity \cite{Fausti2011, Nicoletti2014, Casandruc2015, Nicoletti2016, Nicoletti2018} or metal-insulator transition driven by
electric field \cite{Nakamura2013}.
The development of high precision and controlled computational methods for non-equilibrium
models in strongly correlated regimes is therefore very important.
Even within an approximated framework
such as Dynamical Mean Field Theory \cite{Georges1996, Kotliar2006, Aoki2014} (DMFT),
which reduces bulk lattice problem to the solution of a self-consistent quantum
impurity model, efficient numerically exact real time out-of-equilibrium quantum impurity solver algorithms
are still lacking.

The long time steady state of non-equilibrium strongly interacting quantum systems
is specially difficult to access within high precision numerical methods.
Until recently, most approaches were severely limited
in reaching long times, \textit{e.g.} the density matrix renormalization group (DMRG)\cite{White1992, White1993, Schollwoeck2005}
faces entanglement growth at long times.
Early attempts of real time quantum Monte Carlo \cite{Muehlbacher2008, Werner2009, Werner2010, Schiro2009, Schiro2010}
also experienced an exponential sign problem at long time and large interaction.
Within Monte-Carlo methods, two main routes are currently explored to resolve this issue:
the inchworm algorithm \cite{Cohen2014a, Cohen2014b, Cohen2015, Chen2017a, Chen2017b}
and the so-called ``diagrammatic" Quantum Monte Carlo \cite{Profumo2015} (QMC).
Diagrammatic QMC\cite{Prokofev1998, Mishchenko2000, VanHoucke2008, Prokofev2007,
Prokofev2008, Gull2010, Kozik2010, Pollet2012, VanHoucke2012, Kulagin2013,
Kulagin2013a, Gukelberger2014, Deng2015, Huang2016, Rossi2018a, VanHoucke2019}
computes the perturbative expansion of physical quantities in power of the interaction $U$,
using an importance sampling Monte Carlo.
In Ref.~\onlinecite{Profumo2015}, some of us have shown that, when properly generalized to the Schwinger-Keldysh formalism,
this approach yields the perturbative expansion in the steady state, {\it i.e.} at infinite time.
We showed that, by regrouping the Feynman diagrams into determinants and summing explicitly on the Keldysh indices
of the times of the vertices of the expansion, we eliminate the vacuum diagrams and obtain a clusterisation property allowing us to take the long time limit.
The sum over the Keldysh indices implies a minimal cost of $O(2^n)$ to compute the order $n$,
but {\it uniformly in time}, at any temperature.
We refer to this class of algorithms as ``diagrammatic" for historical reasons,
as their first versions (in imaginary time) were using an explicit Markov chain in the space of Feynman diagrams. However,
modern versions of the algorithms
regroup diagrams with only an exponential number of determinants (instead of sampling the $n!$ diagrams), eliminating the vacuum diagrams,
both in real time\cite{Profumo2015} and in imaginary time\cite{Rossi2017b, Moutenet2018}, which leads to much higher performance.

In this article, we generalize the algorithm presented in Ref.~\onlinecite{Profumo2015} to the
calculations of Green's functions.
Indeed, in its initial formulation it only permits the calculation of physical observables at equal time
such as the density or current, and the full Green's function requires the computation of each time one by one, which is not technically viable.
Here, we show how to use a kernel technique in order to obtain efficiently
the perturbative expansion of the Green's function and the self-energy, as a function of time or frequency.
Computing the Green's function is an important extension of the technique.
First, it is the first step towards building a DMFT real time non-equilibrium impurity solver.
Second, even in the simple context of a quantum dot, each computation provides much more information than the original algorithm
(from which only a single number, {\it e.g.} the current, could be computed).

The central issue of the ``diagrammatic" QMC family is to properly sum the perturbative expansion
away from the weak coupling regime, specially given the fact that one has access to a limited number of orders
(about $10$ in the present case) due to the exponential cost with the order $n$.
Some of us will address this issue in a separate paper\cite{Bertrand2019b}, using the building blocks introduced here.
In this paper, we present the formalism and first benchmark our approach in the weak coupling regime.

This paper is organized as follows: in section \ref{sec:formalism}, we introduce the necessary formalism and derive
the basic equations that will be used to formulate the method. Section
\ref{sec:QMC} discusses our sampling strategy for the QMC algorithm, as well as
relations with previous work.  Section \ref{sec:application} shows our numerical
data and the detailed comparison with our benchmarks in the weak coupling regime.

%%%%%%%%%%%%%%%%%%%%%%%%%%%%%%%%%%%%%%%%%%%%%%%%%%%%%%%%%%%%%%%%%%%%%%%%%%%%%%%
\section{Wick determinant formalism}
\label{sec:formalism}

This section is devoted to the derivation of the main formula needed to set up
the QMC technique. We introduce a systematic formalism that uses what we call
``Wick determinants". Although the formalism is nothing but the usual
diagrammatic expansion (in Keldysh space), its Wick determinant formulation
provides a route for deriving standard results (such as equation of motions) in
a self contained manner that does not require to introduce Feynman diagrams. We
find this approach useful for discussing numerical algorithms.

%%%%%%%%%%%%%%%%%%%%%%%%%%%%%%%%%%%%%%%%%%%%%%%%%%%%%%%%%%%%%%%%%%%%%%%%%%%%%%%
\subsection{Notations and main expansion formula}
We consider a generic class of system given by a time-dependent Hamiltonian of the form,
\begin{eqnarray}
\label{eq:hamiltonian}
\hnd{H}(t) &=&\hnd{H}_0(t)+U \ \hnd{H}_{\rm int}(t) \\
\hnd{H}_0(t) &=& \sum_{xy} \nd{H}^0_{xy}(t)\hnd{c}^\dagger_{x}\hnd{c}_{y} \\
\hnd{H}_{\rm int}(t) &=& \sum_{xy} V_{xy}(t) (\hnd{c}_x^\dagger \hnd{c}_x - \alpha_x) (\hnd{c}_y^\dagger \hnd{c}_y - \alpha_y)
\end{eqnarray}
where $\hnd{H}_0(t)$ is a quadratic unperturbed Hamiltonian. 
The operators $\hnd{c}^\dagger_{x}$ ($\hnd{c}_{x}$) are the usual creation
(destruction) operators on site $x$.
$x$ and $y$ index all discrete degrees of freedom such as sites, orbitals, spin
and/or electron-hole (Nambu) degrees of freedom and will simply be called
orbital indices.
$\hnd{H}_{\rm int}(t)$ is the, possibly time-dependent, electron-electron
interaction perturbation which is assumed to be switched on at $t=0$. Without
loss of generality we assume $V_{xy} = V_{yx}$. We
emphasize that the method described in this paper is not restricted to this
form of interaction (as shown in Ref.~\onlinecite{Profumo2015}) and can be generalized
straightforwardly to arbitrary interactions. However, to improve readability, we
will restrict hereafter our presentation to the case of density-density
interactions. We also add the quadratic shift $\alpha$, which has been
introduced in previous works \cite{Rubtsov2004, Profumo2015, Wu2017}.
In particular, we have shown in Ref.~\onlinecite{Profumo2015} that, in the context of real
time QMC, it can strongly affect
the radius of convergence of the perturbative series. The non-interacting
Hamiltonian is assumed to be already solved, \textit{i.e.} one has calculated all the
one-particle non-interacting Green's functions. Such calculations can be done
even out-of-equilibrium using \textit{e.g.} the formalism discussed in
Ref.~\onlinecite{Gaury2014}.

Our starting point is a formula for the systematic expansion of interacting Green's function in powers of the
parameter $U$. We use the Schwinger-Keldysh formalism to produce this expansion. The Green's functions
acquire additional Keldysh indices $a,a'\in\{0,1\}$ which provides the Green's function with a $2\times 2$ structure,
\be
G^{aa'}_{xx'}(t,t')=
\left(
\begin{array}{cc}
G^T_{xx'}(t,t') & G^<_{xx'}(t,t') \\
G^>_{xx'}(t,t')   &  G^{\bar T}_{xx'}(t,t')
\end{array}
\right)
\ee
where $G^T_{xx'}(t,t')$, $G^<_{xx'}(t,t')$, $G^>_{xx'}(t,t')$ and $G^{\bar T}_{xx'}(t,t')$ are respectively the
standard time ordered, lesser, greater and anti-time ordered Green's functions. We use a similar definition
for the (known) non-interacting Green's function $g^{aa'}_{xx'}(t,t')$. We introduce the Keldysh points $X$ that
gather an orbital index $x$, a time $t$ and a Keldysh index $a$ to form the tuple $X \equiv (x, t, a)$.
Using Keldysh points, we can rewrite the above definitions using the standard conventions of Keldysh formalism,
\begin{equation}
\label{eq:def_G_Keldysh}
G_{xx'}^{aa'}(t,t') 
\equiv -i \left< T_{\rm c} \hnd{c}(x,t,a) \hnd{c}^\dagger(x',t',a') \right>
\end{equation}
where the creation ($\hnd{c}^\dagger(X')$) and annihilation operators
($\hnd{c}(X)$ or $\hnd{c}(x,t,a)$), here in Heisenberg representation, are
ordered using the contour time ordering operator $T_{\rm c}$. $T_{\rm c}$
orders first by Keldysh index ($a$) before ordering by increasing time within
the forward branch ($a=0$), and by decreasing time within the backward branch
($a=1$) eventually multiplying the result by the usual fermionic $(-1)$ factor
whenever an odd number of permutations have been performed. In several places,
we will use the alternative notation for the Green's function
\begin{equation}
G(X, X') \equiv G\left[(x,t,a),(x',t',a')\right]\equiv G_{xx'}^{aa'}(t,t')
\end{equation}
and we will also note the $\delta$ function on the Keldysh contour as 
\begin{equation}
\delta_c\left[((x,t,a),(x',t',a') \right] \equiv \delta(t-t') \delta_{aa'} \delta_{xx'}.
\end{equation}

Using the above notations (with $\hbar = 1$), one can derive the usual
expansion in power of $U$ using Wick's theorem. We first assume
$\alpha_x = 0$ at all orbital indices $x$. We will explain at the end of this paragraph how to extend this formula to the general case $\alpha_x \ne 0$.
We obtain\cite{Profumo2015}:
\begin{multline}\label{eq:def_G_Keldysh_expansion}
G_{xx'}^{aa'}(t,t') = \sum_{n\geq 0}
\frac{i^n U^n}{n!} \int \prod_{k=1}^n du_k
\sum_{\{x_k, y_k\}}\sum_{\{a_k\}}
\times \\
\left(
\prod_{k=1}^n 
(-1)^{a_k} V_{x_k y_k}(u_k)
\right)
\lists{(x,t,a), U_1, \ldots, U_{2n}}{(x',t',a'), U_1, \ldots, U_{2n}}
\end{multline}
which forms the starting point of this work. Here, we have noted for $1\leq k \leq n$ 
\begin{subequations}
\begin{align}
    U_{2k-1} &= (x_k, u_k, a_k) \\
    U_{2k} &= (y_k, u_k, a_k) 
\end{align}
\end{subequations}
and introduced the 
{\it Wick matrix}:
\begin{equation}
\label{eq:def_det}
\lists{A_1, \ldots, A_m}{B_1, \ldots, B_m}_{ij} \equiv
\begin{pmatrix}
    {g}(A_1, B_1) & \hdots & {g}(A_1, B_m) \\
    \vdots & \ddots & \vdots \\
    {g}(A_m, B_1) &  \hdots & {g}(A_m, B_m)
\end{pmatrix}_{ij} 
\end{equation}
where $A_i$ and $B_j$
are two sets of $m$ points on the Keldysh contour. We refer to the determinant of
the Wick matrix as the {\it Wick determinant}. For notation simplicity, the determinant is assumed in the absence of indices,
\begin{equation}
\lists{A_1, \ldots, A_m}{B_1, \ldots, B_m} \equiv
\det \lists{A_1, \ldots, A_m}{B_1, \ldots, B_m}_{ij} 
\end{equation}
In equation (\ref{eq:def_G_Keldysh_expansion}), we start at $t<0$ with a non-interacting state and switch on the interaction for $t\ge 0$.
Hence the time integrals in
Eq.~\ref{eq:def_G_Keldysh_expansion} run over the segment $[0, t_M]$, where $t_M = \max(t,
t')$. The lower boundary simply arises from  $V_{xy}(u<0)=0$.
The upper boundary can be extended to any value larger than $t_M$ without changing the integral's value
(standard property of the Keldysh formalism).
For the practical applications shown in section \ref{sec:QMC}, we will a fixed large value of $t_M$. 
We emphasize that the complexity of the algorithm does {\it not} grow with $t_M$.
Eq.~(\ref{eq:def_G_Keldysh_expansion}) is formally very appealing: it reduces the problem of calculating the contributions of the expansion to a combination of linear algebra and quadrature.

The definition Eq.~(\ref{eq:def_det}) contains an ambiguity that needs to be clarified: the ordering at
equal times of terms like $g(U_{2k},U_{2k})$ is ill defined. For these terms, one must keep the same ordering of the creation and destruction operators as in the
original definition of the interacting Hamiltonian Eq.~(\ref{eq:hamiltonian}), \textit{i.e.} 
\begin{subequations}
    \begin{align}
        g(U_{2k},U_{2k}) &=g^{<}_{y_ky_k}(u_k,u_k)\\
        g(U_{2k-1},U_{2k-1})&=g^{<}_{x_kx_k}(u_k,u_k)\\ 
        g(U_{2k-1},U_{2k})&=g^{>}_{x_ky_k}(u_k,u_k)\\
        g(U_{2k},U_{2k-1})&=g^{<}_{y_kx_k}(u_k,u_k).
    \end{align}
\end{subequations}

To proceed to the general case $\alpha_x \ne 0$, one only needs to shift the diagonal terms of the Wick matrix using the following replacement rules,
as shown in Ref.\onlinecite{Profumo2015}, Appendix A:
\begin{subequations}
    \begin{align}
        g(U_{2k},U_{2k}) &\rightarrow g(U_{2k},U_{2k}) - i\alpha_{y_k} \\
        g(U_{2k-1},U_{2k-1}) &\rightarrow g(U_{2k-1},U_{2k-1}) - i\alpha_{x_k}
    \end{align}
\end{subequations}
As a result, all derivations in this paper can be done by first ignoring
$\alpha_x$, then replacing the non-interacting Green's functions with these
rules. For this reason and for readability, we will keep these replacements
implicit in Wick matrices, but explicit otherwise.

Finally, Eq.~(\ref{eq:def_G_Keldysh_expansion}) can be extended to the
calculations of arbitrary Green's functions. 
The rule for doing so is as follows: whenever one introduces a product
$-i\hnd{c}(Y)\hnd{c}^\dagger(Y')$ under the time ordering operator in
Eq.~(\ref{eq:def_G_Keldysh}), one must add the corresponding Keldysh points in
the Wick determinant of Eq.~(\ref{eq:def_G_Keldysh_expansion}):
\begin{equation}
\lists{X, U_1, \ldots, U_{2n}}{X', U_1, \ldots, U_{2n}}
\rightarrow
\lists{X, Y, U_1, \ldots, U_{2n}}{X', Y', U_1, \ldots, U_{2n}}
\end{equation}

If $Y$ and $Y'$ share the same time and orbital index $y$, we have the possibility to introduce terms of the form
$-i[\hnd{c}(Y)\hnd{c}^\dagger(Y') + \alpha_y]$ in the definition of the Green's
function. In that case, one must replace $g(Y,Y')\rightarrow g(Y,Y') - i\alpha_y$
 in the diagonal of the Wick matrix.
Again, to improve readability, we will keep this replacement implicit
in Wick matrices, but explicit otherwise.

%%%%%%%%%%%%%%%%%%%%%%%%%%%%%%%%%%%%%%%%%%%%%%%%%%%%%%%%%%%%%%%%%%%%%%%%%%%%%%%
\subsection{A few properties of Wick determinants}
Wick determinants have the general properties of determinants: exchanging two Keldysh points on either the
first or the second line of the left hand side of Eq.~(\ref{eq:def_det}) leaves the Wick determinant unchanged up to a factor $(-1)$. 
An important property
of the formalism, as already shown in Ref.~\onlinecite{Profumo2015}, is that for $n>0$:
\begin{equation}
\label{eq:ZisONE}
\sum_{\{a_k\}}
(-1)^{\sum_{k=1}^n a_k} 
\lists{U_1, \ldots, U_{2n}}{U_1, \ldots, U_{2n}} = 0
\end{equation}
for any times $u_1, \ldots, u_n$ and orbital indices $x_1, \ldots, x_n$ and $y_1, \ldots, y_n$.
This relation expresses the fact that vacuum diagrams are automatically cancelled by 
the sum over the Keldysh indices, even before any integration over time.
It is proven in the usual way in the Keldysh formalism, by considering the largest $u_k$ time, say $k=p$.
From the properties of the bare Green's functions, one can show that the elements of the Wick matrix, hence the determinant, 
are in fact all independent of $a_p$ (reflecting the fact that the largest time can be on the upper or the lower part of the contour). 
Therefore the sum over $a_p$ cancels the sum.

We will use the usual expansion of a determinant along one row or one column in terms of
the cofactor matrix. It takes the form
\begin{equation}
\label{eq:det_expand_col}
\lists{A_1,\ldots, A_m}{B_1,\ldots,B_m} = 
\sum_{k=1}^m (-1)^{k+1}{g}(A_k,B_1) \lists{A_1, \ldots \cancel{A_k} \ldots, A_m}{\cancel{B_1}, B_2, \ldots, B_m}
\end{equation}
for the expansion along the first column and
\begin{equation}
\label{eq:det_expand_row}
\lists{A_1,\ldots, A_m}{B_1,\ldots,B_m}= 
\sum_{k=1}^m (-1)^{k+1}{g}(A_1,B_k) \lists{\cancel{A_1}, \ldots  \ldots, A_m}{B_1, \cancel{B_k}, \ldots, B_m}
\end{equation}
for the expansion along the first row. The notation $\cancel{A_k}$ ($\cancel{B_k}$) stands for the fact that the corresponding row or column must be removed from the Wick matrix.

Last, we will also make a systematic use of the fact that the cofactor matrix is directly related to the  inverse of the matrix,
\begin{eqnarray}
\label{eq:inverse} 
 (-1)^{i+j} \lists{A_1\ldots \cancel{A_i}, \ldots  \ldots, A_m}{B_1,\ldots  \cancel{B_j}, \ldots, B_m} =  \nonumber \\
\lists{A_1,\ldots, A_m}{B_1,\ldots,B_m}^{-1}_{ji} 
\lists{A_1,\ldots, A_m}{B_1,\ldots,B_m} 
\end{eqnarray}

%%%%%%%%%%%%%%%%%%%%%%%%%%%%%%%%%%%%%%%%%%%%%%%%%%%%%%%%%%%%%%%%%%%%%%%%%%%%%%%
\subsection{Definition of the kernel $K$ for the one-body Green's function}
\label{sec:kernelG}

In Ref.~\onlinecite{Profumo2015}, a QMC scheme was defined directly on Eq.~(\ref{eq:def_G_Keldysh_expansion}) so that
a single QMC run could provide the value of $G_{xx'}^{aa'}(t,t')$ for a single pair of times $t$ and $t'$.
In order to extend the technique and obtain a full curve (as a function of $t$)
in a single run, a different form must be used. Performing the expansion of
Eq.~(\ref{eq:det_expand_row}) on Eq.~(\ref{eq:def_G_Keldysh_expansion}), we
obtain
\begin{widetext}
\begin{multline}\label{eq:G_Keldysh_expansion2}
G_{xx'}^{aa'}(t,t') = g_{xx'}^{aa'}(t,t') + \sum_{n\geq 1}
\frac{i^n U^n}{n!} \int \prod_{k=1}^n du_k
\sum_{\{x_k, y_k\}}\sum_{\{a_k\}}
\left(
\prod_{k=1}^n 
(-1)^{a_k} V_{x_k y_k}(u_k)
\right)
\times \\
\left(
   \sum_{p=1}^{2n}
(-1)^{p} 
g\left[ (x,t,a), U_p  \right]
\lists{ U_1, \ldots, U_{2n}}{ (x',t',a'), U_1, \ldots, \cancel{U_p},\ldots, U_{2n}}
+
g\left[  (x,t,a), (x',t',a')  \right]
\lists{ U_1, \ldots, U_{2n}}{ U_1, \ldots, U_{2n}}
\right)
\end{multline}
The last term of the sum vanishes for $n>0$ due to Eq.~(\ref{eq:ZisONE}). Factorizing the $g$ from the sum, we arrive at
\begin{equation}\label{eq:G_Keldysh_expansion3}
G_{xx'}^{aa'}(t,t') = g_{xx'}^{aa'}(t,t')  +  
\int du \sum_{b, y} (-1)^b
g_{xy}^{ab}(t,u) K_{yx'}^{ba'}(u,t') 
\end{equation}
where the kernel $K_{yx'}^{ba'}(u,t')=K(Y,X')$ with $Y=(y,u,b)$ is defined by 
\begin{multline}\label{eq:def_kernel_K}
K(Y,X') \equiv (-1)^b
\sum_{n\geq 1}
\frac{i^n U^n}{n!} \int \prod_{k=1}^n du_k
\sum_{\{x_k, y_k\}}\sum_{\{a_k\}}
\left(
\prod_{k=1}^n (-1)^{a_k} V_{x_k y_k}(u_k)
\right)
\sum_{p=1}^{2n}
(-1)^{p} \delta_c(Y, U_p) 
\lists{ U_1, \ldots, U_{2n}}{ X', U_1, \ldots, \cancel{U_p},\ldots, U_{2n}}
\end{multline}
Equations~(\ref{eq:G_Keldysh_expansion3}) and~(\ref{eq:def_kernel_K}) will be the basis of one of the method developed
in this article. Eq.~(\ref{eq:def_kernel_K}) will provide the mean to get a full $t$-curve in a single calculation and
Eq.~(\ref{eq:G_Keldysh_expansion3}) to relate the corresponding kernel to the Green's function $G$, the target of the calculation.

A symmetric kernel $\bar K$ may be derived following the exact same route but now expanding the
Wick determinant along the first {\it column} using  Eq.~(\ref{eq:det_expand_col}). We find
\begin{equation}\label{eq:G_Keldysh_expansion4}
G_{xx'}^{aa'}(t, t') = g_{xx'}^{aa'}(t, t') +  
\int du \sum_{b, y} (-1)^b
\bar K_{xy}^{ab}(t,u) g_{yx'}^{ba'}(u,t') 
\end{equation}
where the kernel $\bar K$ is defined by 
\begin{multline}\label{eq:def_Kernel_Kbar}
   \bar K(X,Y) \equiv (-1)^b
\sum_{n\geq 1}
\frac{i^n U^n}{n!} \int \prod_{k=1}^n du_k
\sum_{\{x_k, y_k\}}\sum_{\{a_k\}}
\left(
\prod_{k=1}^n (-1)^{a_k} V_{x_k y_k}(u_k)
\right)
\sum_{p=1}^{2n}
(-1)^{p} \delta_c(Y,U_p) 
\lists{X, U_1, \ldots, \cancel{U_p},\ldots, U_{2n}}{U_1, \ldots, U_{2n}}
\end{multline}

%%%%%%%%%%%%%%%%%%%%%%%%%%%%%%%%%%%%%%%%%%%%%%%%%%%%%%%%%%%%%%%%%%%%%%%%%%%%%%%
\subsection{Definition of the kernel $L$ of the $F$ Green's function}
\label{sec:kernelF}

Let us define a new Green's function with 4 operators, the $F$ function. As we shall see, the $F$ Green's function can also be
represented in term of a kernel so that we will be able to design a direct QMC method to calculate it. Its interest
stems from the fact that it can be used to reconstruct $G$ while the corresponding QMC technique will be more precise at
high frequency. It is defined as, 
\begin{equation}
\label{eq:def_F_Keldysh}
F_{xx'z}^{aa'}(t,t') 
\equiv (-i)^2 \left< T_{\rm c} \hnd{c}(x,t,a) \hnd{c}^\dagger(x',t',a')\left[\hnd{c}^\dagger(z,t',a')\hnd{c}(z,t',a') -\alpha_z\right] \right>
\end{equation}

In the next paragraph, we shall prove that $F$ is essentially equal to $\bar K$ (up to interacting matrix elements).
The function $F$ is known to provide a better estimate of the Green's function. 
It  has been used in the context of the Numerical Renormalization Group (NRG)\cite{Bulla1998}
as well as in imaginary time QMC methods as an {\it improved estimator} \cite{Hafermann2012}.

The expansion of $F$ reads, 
\begin{equation} \label{eq:F_expand1}
F_{xx'z}^{aa'}(t,t') =
- \sum_{n\geq 0}
\frac{i^{n}U^{n}}{n!} 
\int \prod_{k=1}^n du_k
\sum_{\{x_k, y_k\}}\sum_{\{a_k\}}
\left(
\prod_{k=1}^n 
(-1)^{a_k}
V_{x_k y_k}(u_k)
\right)
\lists{(x,t,a), (z,t',a'), U_1,\ldots, U_{2n}}{ (x',t',a'), (z,t',a'), U_1, \ldots, U_{2n}}
\end{equation}
To obtain the kernel of $F$, we expand the determinant along its first row using Eq.~(\ref{eq:det_expand_row}),
\begin{multline}
F_{xx'z}^{aa'}(t,t') = - \sum_{n\geq 0}
\frac{i^n U^n}{n!} \int \prod_{k=1}^n du_k
\sum_{\{x_k, y_k\}}\sum_{\{a_k\}}
\left(
\prod_{k=1}^n 
(-1)^{a_k} V_{x_k y_k}(u_k)
\right)
\times \\
\left(
   \sum_{p=1}^{2n}
(-1)^{p+1} 
g\Bigl( (x,t,a), U_p  \Bigr)
\lists{ (z,t',a'), U_1, \ldots, U_{2n}}{ (x',t',a'), (z,t',a'), U_1, \ldots, \cancel{U_p},\ldots, U_{2n}}
+ 
\right .
\\
\left .
g\Bigl( (x,t,a), (x',t',a')  \Bigr)
\lists{ (z,t',a'),U_1, \ldots, U_{2n}}{ (z,t',a'),U_1, \ldots, U_{2n}}
-
g\Bigl( (x,t,a), (z,t',a')  \Bigr)
\lists{ (z,t',a'), U_1, \ldots, U_{2n}}{ (x',t',a'), U_1, \ldots, U_{2n}}
\right)
\end{multline}
Identifying the two last terms with the corresponding expansion of $G$, we arrive at,
\begin{equation}\label{eq:F_expand}
F_{xx'z}^{aa'}(t,t') = -g_{xx'}^{aa'}(t,t') [G_{zz}^{<}(t',t') - i\alpha_z] + g_{xz}^{aa'}(t,t') G_{zx'}^{<}(t',t') -
\int du \sum_{b, y} (-1)^b
g_{xy}^{ab}(t,u) L_{yx'z}^{ba'}(u,t')
\end{equation}
where the kernel $L$ is defined by 
\begin{multline}\label{eq:def_Kernel_KF}
L_{yx'z}^{ba'}(u,t') \equiv (-1)^b
\sum_{n\geq 1}
\frac{i^n U^n}{n!} \int \prod_{k=1}^n du_k
\sum_{\{x_k, y_k\}}\sum_{\{a_k\}}
\left(
\prod_{k=1}^n (-1)^{a_k} V_{x_k y_k}(u_k)
\right)
\times \\
\sum_{p=1}^{2n}
(-1)^{p+1} \delta_c\Bigl((y,u,b), U_p\Bigr) 
\lists{ (z,t',a'), U_1, \ldots, U_{2n}}{ (x',t',a'), (z,t',a'), U_1, \ldots, \cancel{U_p},\ldots, U_{2n}}
\end{multline}

%%%%%%%%%%%%%%%%%%%%%%%%%%%%%%%%%%%%%%%%%%%%%%%%%%%%%%%%%%%%%%%%%%%%%%%%%%%%%%%
\subsection{Relation between $F$, $\bar K$ and $G$: equations of motion}
\label{sec:eq_of_motion}
Here, we show that the expressions for the different kernels can
be formally integrated to provide connections between the different kernels and
Green's functions. We will arrive at expressions that are essentially what can be obtained
directly using equations of motions. 

We start with the expression of $\bar K$, Eq.~(\ref{eq:def_Kernel_Kbar}).
The first step is to realize that the sum over the $2n$ determinants provide identical contributions
to the kernel. Indeed, odd $p=2k-1$ and even $p=2k$ values of $p$ provide identical contributions due to the symmetry of
$V_{xy}$. Similarly, odd values $p=2k-1$ have the same contribution as $p=1$ as can be shown by using the symmetry properties 
of the determinant and exchanging the role of $U_1 \leftrightarrow U_{2k-1}$ and $U_2 \leftrightarrow U_{2k}$ in the sums 
and integration. We arrive at,
\begin{multline}
   \bar K_{xy}^{ab}(t,u) = (-1)^b
\sum_{n\geq 1}
\frac{i^n U^{n}}{n!} 
\int d u_1 \sum_{x_1, y_1} \sum_{a_1} (-1)^{a_1} V_{x_1, y_1}(u_1)
\int \prod_{k=2}^n du_k
\sum_{\{x_k, y_k\} \atop k\ge 2}\sum_{\{a_k\} \atop k\ge 2}
\left(
\prod_{k=2}^n (-1)^{a_k} V_{x_k y_k}(u_k)
\right)
\times \\
2n \,
\delta_c(U_1, (y,u,b))
\lists{(x,t,a), U_2, U_3,\ldots, U_{2n}}{ U_1, U_2, U_3, \ldots,  U_{2n}}
\end{multline}
We can now perform explicitly the integral over $u_1$ where the delta function yields, for $u \in [0, t_M]$ ($\bar K$ is zero otherwise):
\begin{align}
   \bar K_{xy}^{ab}(t,u) &=  2iU
\sum_{n\geq 1}
\frac{i^{n-1}U^{n-1}}{(n-1)!} 
\sum_{y_1}  V_{y, y_1}(u)
\int \prod_{k=2}^n du_k
\sum_{\{x_k, y_k\}\atop {\{a_k\} \atop {k \ge 2}}}
\left(
\prod_{k=2}^n 
(-1)^{a_k}
V_{x_k y_k}(u_k)
\right)
\lists{(x,t,a), (y_1,u,b), U_3, \ldots, U_{2n}}{ (y,u,b), (y_1,u,b), U_3, \ldots, U_{2n}}
\\
\bar K_{xy}^{ab}(t,u) &=  2iU
\sum_{z} V_{y, z}(u)
\sum_{n\geq 0}
\frac{i^{n}U^{n}}{n!} 
\int \prod_{k=1}^n du_k
\sum_{\{x_k, y_k\}\atop \{a_k\} }
\left(
\prod_{k=1}^n 
(-1)^{a_k}
V_{x_k y_k}(u_k)
\right)
\lists{(x,t,a), (z,u,b), U_1,\ldots, U_{2n}}{ (y,u,b), (z,u,b), U_1, \ldots, U_{2n}}
\\
\label{eq:relation_Kbar_F}
\bar K_{xy}^{ab}(t,u) &= - 2iU \sum_z V_{yz}(u) F^{ab}_{xyz}(t,u)
\end{align}
This shows the kernel $\bar K$ is no more than a sum of 2-particle Green's
functions. The relation between $\bar K$ and $G$
in Eq.~(\ref{eq:G_Keldysh_expansion4}) can then be transformed into:
\begin{equation}\label{eq:relation_F_G}
G_{xx'}^{aa'}(t, t') = g_{xx'}^{aa'}(t, t') -
2 i U  \int du \sum_{b, y} (-1)^b
\sum_{z}  V_{yz}(u) F_{xyz}^{ab}(t,u) g_{yx'}^{ba'}(u,t') 
\end{equation}
which can be used to reconstruct $G$ from the knowledge of $F$. This relation
is the well known equation of motion for $G$. It also shows that
$F$ is essentially the convolution of $G$ with the self-energy.

As a side note, we show in Appendix~\ref{app:L_as_gf} that the kernel $L$ can
also be expressed in terms of Green's functions by following the same formalism, in accordance with the
equation of motion for $F$.

%%%%%%%%%%%%%%%%%%%%%%%%%%%%%%%%%%%%%%%%%%%%%%%%%%%%%%%%%%%%%%%%%%%%%%%%%%%%%%%
\subsection{Retarded and advanced kernels}

As the retarded (or advanced) Green's functions directly give the spectral
functions, they are of particular interest.  At equilibrium, they contain all
information which can be obtained from the Keldysh Green's function.  We show
here simple relations to compute them from the kernels $K$, $\bar K$ or $L$.

The retarded and advanced Green's functions relate to the lesser and greater Green's functions as follows:
\begin{align}
    \label{eq:def_retarded_1}
    &G^R_{xx'}(t,t') = \theta(t-t') \left( G^>_{xx'}(t,t') - G^<_{xx'}(t,t') \right) \\
    \label{eq:def_advanced_1}
    &G^A_{xx'}(t,t') = \theta(t'-t) \left( G^<_{xx'}(t,t') - G^>_{xx'}(t,t') \right)
\end{align}
where $\theta$ is the Heaviside function.
From the definitions of the time-ordered and time-antiordered Green's functions, these can also be written as:
\begin{align}
    \label{eq:def_retarded_2}
    &G^R_{xx'}(t,t') = G^{a0}_{xx'}(t,t') - G^{a1}_{xx'}(t,t') \\
    \label{eq:def_advanced_2}
    &G^A_{xx'}(t,t') = G^{0a}_{xx'}(t,t') - G^{1a}_{xx'}(t,t')
\end{align}
where $a$ can be any Keldysh index. These are also valid for the non-interacting $g$.
As $\bar K$ is a sum of Green's function, one may define in the same way a
retarded version of $\bar K$, denoted $\bar{K}^R$. We can see from
Eq.~(\ref{eq:relation_Kbar_F}) and the definition of $F$ in
Eq.~(\ref{eq:def_F_Keldysh}) that $\bar{K}^R$ follows the same properties:
\begin{equation}
    \bar{K}^R_{xx'}(t,t') = \bar{K}^{a0}_{xx'}(t,t') - \bar{K}^{a1}_{xx'}(t,t')
\end{equation}

We now show a simple relation between $G^R$, $g^R$ and $\bar{K}^R$. Plugging Eq.~(\ref{eq:G_Keldysh_expansion4}) into Eq.~(\ref{eq:def_retarded_2}), one gets:
\begin{equation}
    G^R_{xx'}(t,t') = g^R_{xx'}(t,t')
    + \int du \sum_y \left(
        \bar{K}^{00}_{xy}(t,u) [ g^{00}_{yx'}(u,t') - g^{01}_{yx'}(u,t') ]
        - \bar{K}^{01}_{xy}(t,u) [ g^{10}_{yx'}(u,t') - g^{11}_{yx'}(u,t') ]
    \right)
\end{equation}
This simplifies into:
\begin{equation}
    G^R_{xx'}(t,t') = g^R_{xx'}(t,t')
    + \int du \sum_y
    \bar{K}^R_{xy}(t,u) g^R_{yx'}(u,t')
\end{equation}

Similar relations may be derived with $K$, $F$ and $L$.
In fact, for any function from $K$, $F$ and $L$, which all depends on two times and two Keldysh indices, we choose to define a retarded and advanced function in the same way as Eq.~(\ref{eq:def_retarded_1}) and Eq.~(\ref{eq:def_advanced_1}).
As all of them are sums of Green's functions, one may show that they all verify similar properties as in Eq.~(\ref{eq:def_retarded_2}) and~(\ref{eq:def_advanced_2}).
Then from Eq.~(\ref{eq:G_Keldysh_expansion3}) follows:
\begin{equation}
    \label{eq:GA_from_KA}
    G^A_{xx'}(t,t') = g^A_{xx'}(t,t')
    + \int du \sum_y
    g^A_{xy}(t,u) K^A_{yx'}(u,t')
\end{equation}
and from Eq.~(\ref{eq:F_expand}) follows:
\begin{equation}
    F^A_{xx'z}(t,t') = -g^A_{xx'}(t,t')[G^<_{zz}(t',t') - i\alpha_z]
    + g^A_{xz}(t,t')G^<_{zx'}(t',t)
    - \int du \sum_y g^A_{xy}(t,u) L^A_{yx'z}(u,t')
\end{equation}

%%%%%%%%%%%%%%%%%%%%%%%%%%%%%%%%%%%%%%%%%%%%%%%%%%%%%%%%%%%%%%%%%%%%%%%%%%%%%%%
\section{Quantum Monte Carlo technique}
\label{sec:QMC}

We now turn to the stochastic algorithms that will be used for the actual evaluations of the multi-dimensional
integrals that define the expansion of the Green's function. These algorithms are direct extensions of
the algorithm of Ref.~\onlinecite{Profumo2015} and inherit of most of its properties. The main novelty lies in using kernels which permits the calculation of the full time dependency of the Green's function in a single QMC run.

%%%%%%%%%%%%%%%%%%%%%%%%%%%%%%%%%%%%%%%%%%%%%%%%%%%%%%%%%%%%%%%%%%%%%%%%%%%%%%%
\subsection{Sampling of the kernel $K$}
Let us first discuss the calculation of $G$ using the kernel $K$, using Eq.~(\ref{eq:G_Keldysh_expansion3}) and~(\ref{eq:def_kernel_K}).
We rewrite by explicitly separating the sum over Keldysh indices (which will be summed explicitly) and the sum over space and integral over time
(which will be sampled using Monte-Carlo). This separation was shown to be extremely important in Ref.~\onlinecite{Profumo2015}. The kernel takes the form, 
\begin{align}\label{eq:kernelQMC1}
K(Y,X') &= (-1)^b
\sum_{n\geq 1}
\int \prod_{k=1}^n du_k
\sum_{\{x_k, y_k\}}
\sum_{p=1}^{2n}
\sum_{a_p}
(-1)^{a_p} 
\delta_c\Bigl(Y, U_p\Bigr) 
W^n_p(X',\{U_k\}, a_p)
\\
W^n_p(X',\{U_k\}, a_p)
&\equiv  
\frac{i^n U^n}{n!} 
\left(\prod_{k=1}^n  V_{x_k y_k}(u_k) \right)
\sum_{ \{a_k\} \atop k\neq p}
\left( \prod_{k\neq p} (-1)^{a_k} \right) 
(-1)^p
\lists{ U_1, \ldots, U_{2n}}{ (x',t',a'), U_1, \ldots, \cancel{U_p},\ldots, U_{2n}}
\end{align}
We define a configuration $\cal C$ as 
\begin{itemize}
   \item The order $n$.
   \item A set of times $\{u_k \in [0, t_M]\}$ for $ 1\leq k \leq n$.
   \item Two sets of indices $\{x_k\}$ and $\{y_k\}$ for $ 1\leq k \leq n$.
\end{itemize}
and the sum over all configuration as the integral over the $u_k$ and the sum over the $x_k,y_k$.
For practical purpose, the time integrals run over a finite interval $[0,
t_M]$. In accordance with the remark following
Eq.~(\ref{eq:def_G_Keldysh_expansion}), $t_M$ can be chosen to be any value
larger than $t$ and $t'$ of the target Green's functions $G(t, t')$.

The kernel takes the form,
\begin{equation}\label{eq:kernelQMC2}
K(Y,X') = (-1)^b
\sum_{{\cal C}}
\left (
\sum_{p=1}^{2n}
\sum_{a_p}
   (-1)^{a_p}  \delta_c\Bigl(Y, U_p\Bigr) 
   W^n_p(X', \{U_k\}, a_p)  
\right)
\end{equation}
where the sum over the configurations $\cal C$ is a compact notation for the sum and integrals of Eq.\eqref{eq:kernelQMC1}.
We observe that a single configuration provides values of $K$ for $2n$ different points $Y$ through the delta function in the preceding expression.
To sample the sum over configurations, we need to define a positive function $W(\cal C)$ that will provide the (unnormalized) probability for the configuration $\cal C$
to be visited by the QMC algorithm. We define this weight as,
\begin{equation}
    W({\cal C}) = \sum_{p=1}^{2n} \sum_{a=0,1}  \left | W^n_p(X',\{U_k\}, a) \right |
\end{equation}
Noting $Z_{\rm QMC} \equiv \sum_{\cal C} W(\cal C)$, the kernel can be rewritten as,
\begin{equation}\label{eq:kernelQMC3}
    K(Y,X') = (-1)^b Z_{\rm QMC}
\left\langle
\sum_{p=1}^{2n}
\sum_{a_p}
   (-1)^{a_p} 
   \dfrac{W^n_p(X', \{U_k\}, a_p)}{W({\cal C})}
   \delta_c\Bigl((y,u,b), U_p\Bigr) 
\right\rangle
\end{equation}
where the average is taken over the distribution $W({\cal C})/Z_{\rm QMC}$.
$Z_{\rm QMC}$ is an effective partition function associated to the QMC algorithm.
Note that by construction, the weight $W({\cal C})$ {\it controls} the
measurement, {\it i.e.} $|W^n_p(X',\{U_k\}, a)| \leq W({\cal C})$. This is an
essential property for a reweighting technique since it guarantees that the
weight $W^n_p(X',\{U_k\}, a)/W({\cal C})$ does not diverge (which can produce
an infinite variance, for an example of this effect in the context of
determinantal Monte-Carlo see \textit{e.g.} Ref.~\onlinecite{Shi2016}).

To sample $W({\cal C})$ and evaluate $Z_{\rm QMC}$, we use the continuous time
Monte-Carlo technique that was discussed in details in Ref.~\onlinecite{Profumo2015}.
We use moves that change the order $n$ by $\pm 1$ so that all orders (up to a
maximum one) can be calculated in a single run. The algorithm has very good
ergodicity properties since the order $n=0$ configuration is visited regularly.
Each configuration $\cal C$ provides $2n$ values of $Y=(y,u,b)$ which are
recorded by binning with the weight of Eq.~(\ref{eq:kernelQMC3}).

The partial weights $W_p^n$ possess an essential {\it clusterization property},
which generalizes the one discovered in Ref.~\onlinecite{Profumo2015}: if one or
several times $u_k$ goes to infinity ({\it i.e.} is far from
$t'$), then all $W_p^n$ goes to 0. In other words the integrand is localized
around $t'$. A detailed proof is provided in Appendix~
\ref{app:ClusterizationWProof}. The point $X'= (x',t',a')$ is kept fixed during
the calculation to {\it anchor} the integral around this point.
An important consequence of the clusterization property is that the computational cost of the
algorithm is uniform in $t_M$. Indeed, as one increases $t_M$, one simply adds regions of the configuration space that have a vanishingly small weight, hence do not contribute to the integral and do not get sampled.

Last, in order to calculate the factors $W_p^n$, we use the fact that they are
made of the cofactors of the original matrix, hence can be rewritten as,
\begin{align}\label{eq:kernelQMC4}
    W^n_p(X',\{U_k\}, a_p)
&=
-\frac{i^n U^n}{n!} 
\left(\prod_{k=1}^n  V_{x_k y_k}(u_k)\right)
\sum_{ \{a_k\} \atop k\neq p}
\left( \prod_{k\neq p} (-1)^{a_k} \right) 
\lists{ X, U_1, \ldots, U_{2n}}{ X', U_1 ,\ldots, U_{2n}}_{p1}^{-1} \times
\lists{ X, U_1, \ldots, U_{2n}}{ X', U_1 ,\ldots, U_{2n}}
\end{align}
This last form is very convenient since a single Wick matrix (and its inverse) need to be stored and monitored during the calculation.

%%%%%%%%%%%%%%%%%%%%%%%%%%%%%%%%%%%%%%%%%%%%%%%%%%%%%%%%%%%%%%%%%%%%%%%%%%%%%%%
\subsection{Sampling of the kernel $L$}

Following the same route for $L$ as was done for $K$, we can write:
\begin{align}
    L_{yx'z}^{ba'}(u,t') &= (-1)^b
\sum_{n\geq 1}
\int \prod_{k=1}^n du_k
\sum_{\{x_k, y_k\}}
\sum_{p=1}^{2n}
\sum_{a_p} (-1)^{a_p}
\delta_c\Bigl((y,u,b), U_p\Bigr)
W_{p+2}^n(X', \{U_k\}, a_p, z) \\
W_{p+2}^n(X', \{U_k\}, a_p, z)
&\equiv
\frac{i^n U^n}{n!}
\left(\prod_{k=1}^n  V_{x_k y_k}(u_k) \right)
\sum_{ \{a_k\} \atop k\neq p}
\left( \prod_{k\neq p} (-1)^{a_k} \right)
(-1)^{p+1}
\lists{ (z,t',a'), U_1, \ldots, U_{2n}}{ (x',t',a'), (z,t',a'), U_1, \ldots, \cancel{U_p},\ldots, U_{2n}}
\end{align}
thus defining $W_{p}^n$ for $p = 3, 4, \ldots, 2n+2$. We also define $W_1^n$ and $W_2^n$ in the following way:
\begin{align}
    W_{1}^n(X', \{U_k\}, z)
&\equiv
-\frac{i^n U^n}{n!}
\sum_{ \{a_k\}}
\left( \prod_{k=1}^n (-1)^{a_k}
V_{x_k y_k}(u_k)  \right)
\lists{ (z,t',a'), U_1, \ldots, U_{2n}}{(z,t',a'), U_1, \ldots, U_{2n}}
\\
W_{2}^n(X', \{U_k\}, z)
&\equiv
\frac{i^n U^n}{n!}
\sum_{ \{a_k\}}
\left( \prod_{k=1}^n (-1)^{a_k}
V_{x_k y_k}(u_k)  \right)
\lists{ (z,t',a'), U_1, \ldots, U_{2n}}{(x',t',a'), U_1, \ldots, U_{2n}}
\end{align}
These two extra values are necessary to compute $G^<_{zz}(t',t')$ and $G^<_{zx'}(t',t')$, which are needed to obtain $F$, as can be seen in Eq.~(\ref{eq:F_expand}).
Moreover, they do not require extra computation time, as they are a direct by-product of the computation of the $W_p^n$ for $p > 2$.
Indeed, in the same spirit as Eq.~(\ref{eq:kernelQMC4}), the determinant within any $W_p^n$ (for $p = 1, \ldots, 2n+2$) can be replaced by:
\begin{equation}
(-1)^{p+1}
\lists{ X, (z,t',a'), U_1, \ldots, U_{2n}}{ X', (z,t',a'), U_1 ,\ldots, U_{2n}}_{p1}^{-1} \times
\lists{ X, (z,t',a'), U_1, \ldots, U_{2n}}{ X', (z,t',a'), U_1 ,\ldots, U_{2n}}
\end{equation}
Again, a single Wick matrix is needed to get contributions to all $W_p^n$ (given a set of Keldysh indices), which is very convenient.

Configurations are defined in the same way as in the previous section, and the weight of a configuration $\cal C$ is now:
\begin{equation}
    W({\cal C}) = \left| W^n_1(X',\{U_k\}, z) \right| 
    + \left| W^n_2(X',\{U_k\}, z) \right| 
    + \sum_{p=1}^{2n} \sum_{a=0,1}  \left| W^n_{p+2}(X',\{U_k\}, a, z) \right|
\end{equation}
We define again $Z_{\rm QMC} \equiv \sum_{\cal C} W(\cal C)$ (which however has a different value than in the previous section).
Finally, $L$ can be written as:
\begin{equation}
L_{yx'z}^{ba'}(u,t') = (-1)^b Z_{\rm QMC}
\left\langle
\sum_{p=1}^{2n}
\sum_{a_p}
   (-1)^{a_p}
   \dfrac{W^n_{p+2}(X', \{U_k\}, a_p, z)}{W({\cal C})}
   \delta_c\Bigl((y,u,b), U_p\Bigr)
\right\rangle
\end{equation}
and, from Eq.~(\ref{eq:def_G_Keldysh_expansion}), we get that the values of $G^<_{zz}(t',t')$ and $G^<_{zx'}(t',t')$ (needed to compute $F$) are:
\begin{align}
    G^<_{zz}(t',t') =&
    - Z_{\rm QMC}
    \left< \dfrac{W_1^n(X', \{U_k\}, z)}{W({\cal C})} \right>
    \\
    G^<_{zx'}(t',t') =&
    Z_{\rm QMC}
    \left< \dfrac{W_2^n(X', \{U_k\}, z)}{W({\cal C})} \right>
\end{align}
The Monte-Carlo algorithm used to evaluate these averages is the same as in the previous section, except for the weight $W({\cal C})$.

\end{widetext}

%%%%%%%%%%%%%%%%%%%%%%%%%%%%%%%%%%%%%%%%%%%%%%%%%%%%%%%%%%%%%%%%%%%%%%%%%%%%%%%
\subsection{A discussion of the Werner \textit{et al.} approach\cite{Werner2010}}

\begin{figure}[t]
    \centering
    \includegraphics[width=7cm]{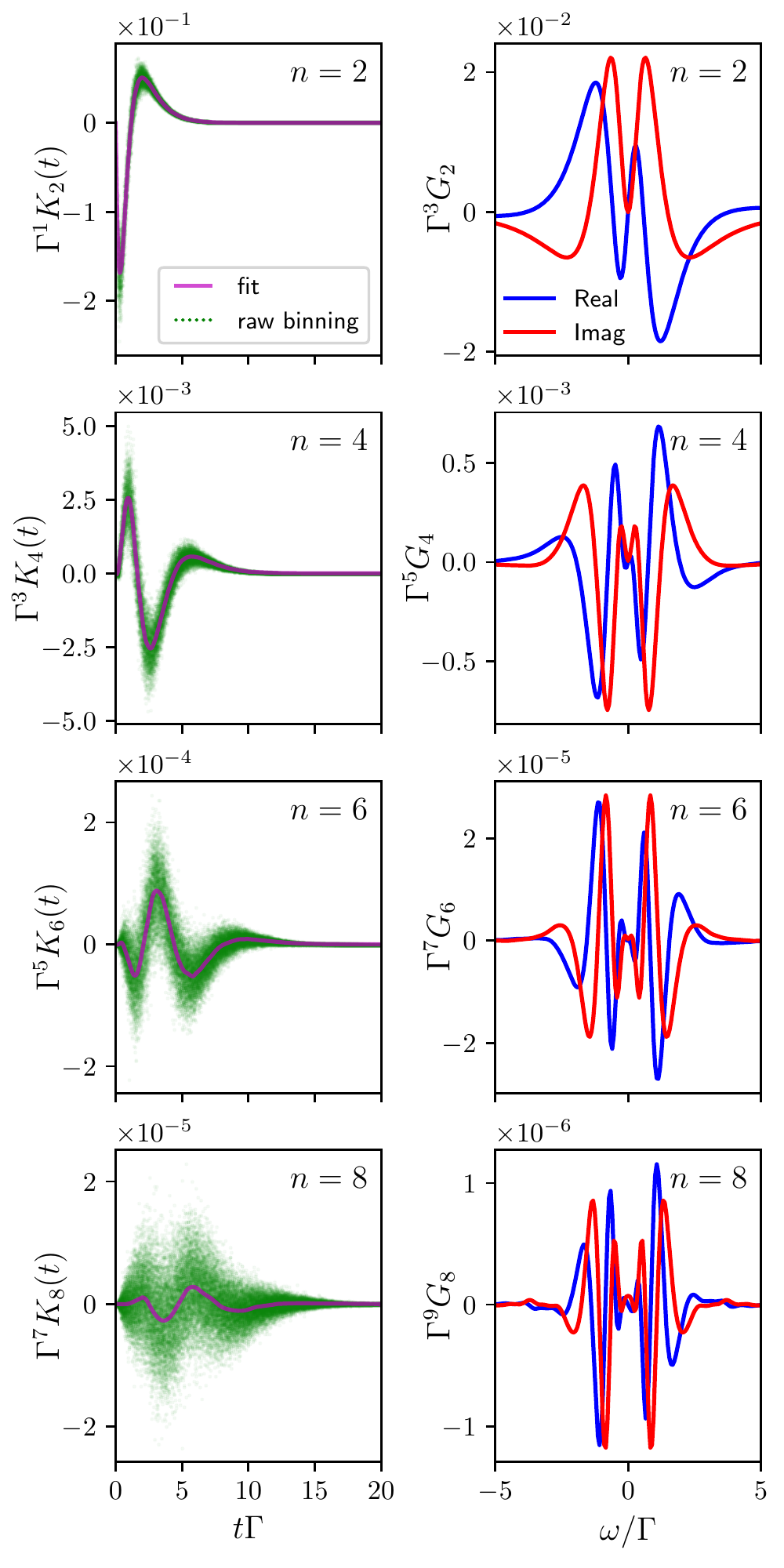}
    \caption{
        \label{fig:kernels}
        First non-zero orders of the retarded kernel in time (left column), and
        corresponding retarded Green's function in frequency (right column),
        for the particle-hole symmetric model $\epsilon_d=0$. The green dots in the
        left column correspond to the raw data of the binning with apparent
        noise arising from high frequencies.  The purple lines are a fit of the
        kernel, shown for illustration purpose only, where the high frequency
        noise has been subtracted by smearing the cumulative function of the
        kernel. Maximum time is $t_M = 20/\Gamma$. One can see (lower left
        panel) that at order $n=8$ a lower integration time would not capture
        the whole kernel, thus the steady state would not be reached.
    }
\end{figure}

In this paragraph, we discuss the relation of this work with a preceding 
work \cite{Werner2009,Werner2010} that also implements an expansion in powers of $U$ within the Keldysh formalism. 
Although both results are consistent, Ref.~\onlinecite{Werner2010} has two important limitations which are not present in the method presented here. 
First, it suffers from a very large sign problem that increases drastically
with time, while we do not experience a sign problem. 
Typical data shown in Ref.~\onlinecite{Werner2010} corresponds to a maximum time of
$t_M\approx 5/\Gamma$ between the switching of the interaction and the
measurement of the observable while we found that $t_M\approx 20/\Gamma$ is
needed to enter the stationary result at order $n=8$ as shown in
Fig.~\ref{fig:kernels}. 
A direct consequence of this issue is that Ref.~\onlinecite{Werner2010} cannot access
the small bias regime where the Kondo effect is present: since the Kondo
temperature $T_K$ is typically much smaller than $\Gamma$, long simulation
times $t_M \gg 1/T_K$ are needed to capture the Kondo physics. 
Second, the technique of Ref.~\onlinecite{Werner2010} suffered from a very large sign
problem outside of the electron-hole symmetry point so that only this point
could be studied. 

An interesting aspect of Ref.~\onlinecite{Werner2010} is that some results could be
obtained in regimes where the ``sign" of the Monte-Carlo calculation was very
small $\sim 10^{-3}$ (see for instance Fig.5 of Ref.~\onlinecite{Werner2009}). Such a
small sign is usually associated with very large error bars that prevents
practical calculations to be performed. In the rest of this paragraph, we 
make a simple technical remark that explains the origin of this
behaviour.

The main expansion formula used in the present work is Eq.~(\ref{eq:def_G_Keldysh_expansion}) which provides the expansion for the Keldysh Green's function $G_{xx'}^{aa'}(t,t')$.
A similar formula\cite{Profumo2015} provides the sum of vacuum diagrams, sometimes called the Keldysh ``partition function" $Z$,
\begin{multline}
\label{eq:Z}
Z \equiv \sum_{n\geq 0}
\frac{i^n U^n}{n!} \int \prod_{k=1}^n du_k
\sum_{\{x_k, y_k\}}
\left(
\prod_{k=1}^n  V_{x_k y_k}(u_k)
\right)
\times \\
\sum_{\{a_k\}}
\prod_{k=1}^n (-1)^{a_k} 
\lists{ U_1, \ldots, U_{2n}}{U_1, \ldots, U_{2n}}
\end{multline}
We have $Z=1$ in the Keldysh formalism, reflecting the unitarity of  quantum mechanics.
We see that (\ref{eq:ZisONE}) indeed implies $Z=1$, and that the cancellation of  vacuum diagrams is due to the sum over Keldysh indices.
The integrand is identically zero.

Let's note $Z(\{U_i\})$ the integrand of Eq.~(\ref{eq:Z}) (without the sum over Keldysh indices):
\begin{equation}
Z(\{U_i\}) \equiv 
\frac{i^n U^n}{n!}
\left(
\prod_{k=1}^n (-1)^{a_k} V_{x_k y_k}(u_k)
\right)
\lists{ U_1, \ldots, U_{2n}}{U_1, \ldots, U_{2n}}
\end{equation}
Ref.~\onlinecite{Werner2009} Monte-Carlo samples the absolute value of this integrand $|Z(\{U_i\})|$
(the authors actually used auxiliary Ising variables
but that does not impact the present argument). We also note $G(X,X', \{U_i\})$
the integrand of Eq.~(\ref{eq:def_G_Keldysh_expansion}) and $Z_{\rm QMC}$ the
integral of $|Z(\{U_i\})|$:
\begin{equation}
        Z_{\rm QMC} \equiv \sum_{n\geq 0}
 \int \prod_{k=1}^n du_k
\sum_{\{x_k, y_k\}}\sum_{\{a_k\}}
|Z(\{U_i\})|
\end{equation}
The weight of the QMC is $ |Z(\{U_i\})|/Z_{\rm QMC}$. We have 
\begin{equation}
\label{eq:werner}
G_{xx'}^{aa'}(t,t') = \frac{\left\langle\frac{ G(X,X',\{U_i\})}{|Z(\{U_i\})|}\right\rangle}
{\left\langle \frac{Z(\{U_i\})}{|Z(\{U_i\})|}\right\rangle}
\end{equation}
where the average is the Monte-Carlo average.
The denominator of Eq.~(\ref{eq:werner}) is the QMC sign mentioned above. From
$Z=1$, we find that this QMC sign is simply given by 
$\left\langle Z(\{U_i\})/|Z(\{U_i\})|\right\rangle = 1/Z_{\rm QMC}$.

Now, we note that the probability to visit order 0 is also $|Z(\varnothing)|/Z_{\rm QMC} = 1/Z_{\rm QMC}$. 
Therefore the average sign in the denominator of Eq.~(\ref{eq:werner}) is the
probability to visit a configuration at zeroth order. This probability
decreases when $U$ is increased, or at long time, when higher orders are sampled which explains
why the QMC sign was observed to decrease drastically in Ref.~\onlinecite{Werner2009}.
However, this QMC sign is always positive and could a priori be computed very
efficiently as an integral of a positive function using \textit{e.g.} the
technique in Ref.~\onlinecite{Profumo2015}.
A genuine sign problem can however
results from the numerator of Eq.~(\ref{eq:werner}).

\begin{figure}[h]
    \centering
    \includegraphics[width=5cm]{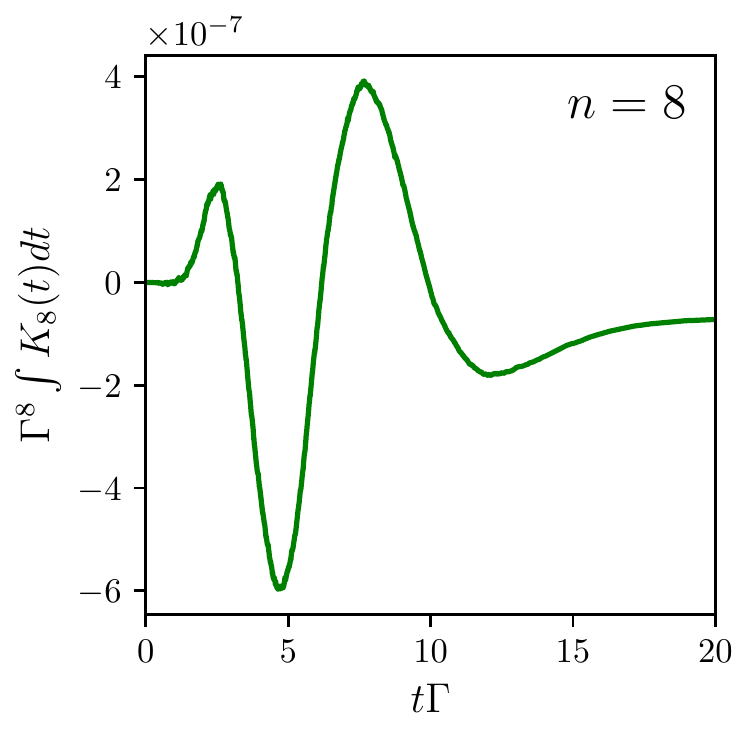}
    \caption{
        \label{fig:primitive_kernel}
        Cumulative function $\int_0^t K^R_8(u)du$ obtained by integrating the
        raw data of the lower left panel of Fig.~\ref{fig:kernels}. Taking the
        integral gets rid of the apparent noise of the raw data whose origin is
        simply the presence of the binning grid in time.
    }
\end{figure}

\begin{figure*}[t]
    \centering
    \includegraphics[width=\textwidth]{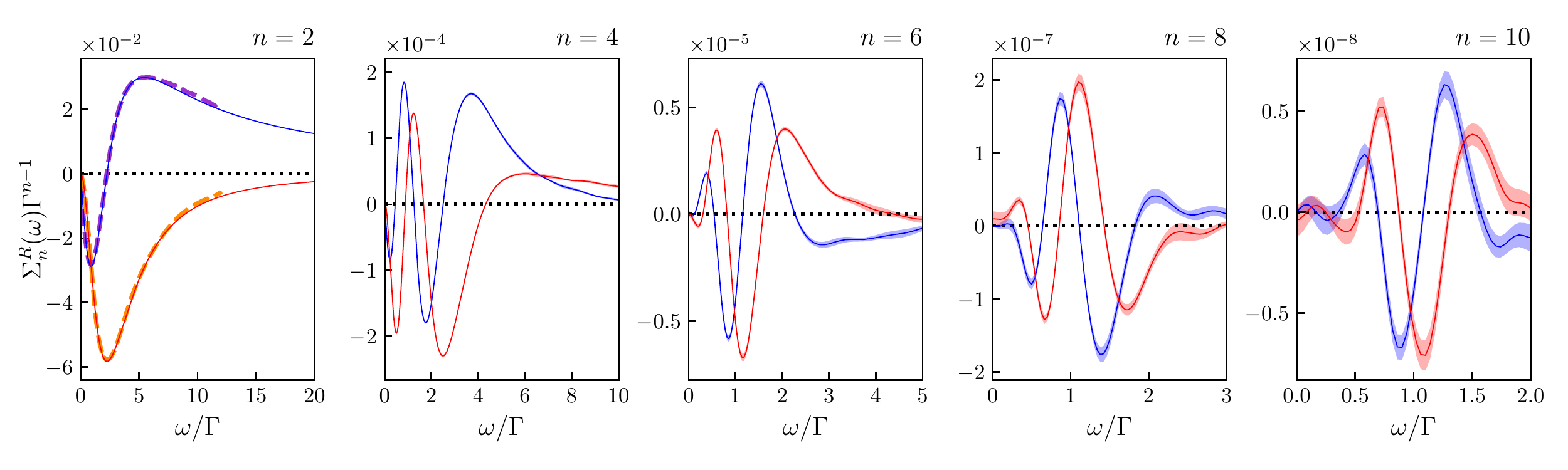}
    \caption{
        \label{fig:SE_order_by_order}
        First terms $\Sigma^R_n(\omega)$ of the development of the retarded self-energy in the
         particle-hole symmetric case ($\epsilon_d=0$) for $n=2,4,6,8$ and $10$.
        (plain lines, real part in blue and imaginary part in red). 
        This curves are obtained in a single Monte-Carlo run. 
        Error bars are shown as shaded areas.  
        A previous result at order two from Yamada\cite{Yamada1975} is shown in dashed lines. 
        Note the decreasing scale with $n$.
        Maximum integration time is $t_M = 20/\Gamma$.
    }
\end{figure*}

%%%%%%%%%%%%%%%%%%%%%%%%%%%%%%%%%%%%%%%%%%%%%%%%%%%%%%%%%%%%%%%%%%%%%%%%%%%%%%%%%%%
\section{Application to the Anderson impurity model}
\label{sec:application}

We now turn to the illustration of our new techniques with calculations done on the Anderson impurity model.
The implementation of our technique was based on the TRIQS package\cite{Parcollet2015}.
We only present here results that showcase the technique and differ the exploration of the physics of the model,
in particular the Kondo physics out of equilibrium, to the companion article of the present work \cite{Bertrand2019b}.
We stress that the QMC technique
is not restricted to impurity models and also applies to lattice models such as the Hubbard model.  

%%%%%%%%%%%%%%%%%%%%%%%%%%%%%%%%%%%%%%%%%%%%%%%%%%%%%%%%%%%%%%%%%%%%%%%%%%%%%%%
\subsection{Definition of the model}
In the Anderson impurity model, the impurity is described by the operators $\hnd{d}^\dagger_{\sigma}$ ($\hnd{d}_{\sigma}$) that create (destroy) 
an electron on the impurity with spin $\sigma$. The impurity is connected to two non-interacting electrodes via a tunneling Hamiltonian. Instead
of providing explicitly this tunneling Hamiltonian and the Hamiltonian of the electrodes, it is simpler to write directly the non-interacting Green's function of the
impurity. We work with its wide band form which is appropriate for the low
energy physics of a regular impurity model. The retarded Green's function reads
in the frequency domain,
\begin{equation}
\label{eq:aim_retarded}
g^R_{\sigma\sigma'}(\omega) = \frac{1}{\omega - \epsilon_d + i\Gamma} \delta_{\sigma\sigma'}
\end{equation}
where the parameter $\Gamma$ sets the width of the resonance in the absence of
electron-electron interactions and the on-site energy $\epsilon_d$ sets the
resonance with respect to the Fermi level. Eq.~(\ref{eq:aim_retarded}) fully
defines the model at equilibrium. The model is made non trivial through the
interacting terms that reads,
\begin{eqnarray}
\hnd{H}_{\rm int}
= U\theta(t)\left(\hnd{n}_\uparrow-\frac{1}{2}\right)\left(\hnd{n}_\downarrow-\frac{1}{2}\right).
\end{eqnarray}
where $\hnd{n}_\sigma = \hnd{d}^\dagger_{\sigma} \hnd{d}_{\sigma}$ is the impurity electronic density of spin $\sigma$
and the Heaviside function $\theta(t)$ represents the fact that the interaction is switched on at $t=0$.
The calculations are performed up to large times $t_M$ so that the system has relaxed to its stationary regime which corresponds to the interacting system at the bath temperature.
All calculations are performed at very low temperature $k_{\rm B}T =
10^{-4}\Gamma$, although the method is suitable for finite temperature as well.

The main output of our calculations is the expansion for the interacting retarded Green's function. Restricting ourselves to the stationary limit,
it is a function of $t-t'$ only and can be studied in the frequency domain. 
\be
G^R_{\sigma\sigma'}(t - t') = \delta_{\sigma\sigma'} \sum_{n=0}^{+\infty} G_n(t-t') U^n
\ee
from which one can obtain the corresponding quantity in the frequency domain by fast Fourier transform,
\be
G^R_{\sigma\sigma'}(\omega) = \delta_{\sigma\sigma'} \sum_{n=0}^{+\infty} G_n(\omega) U^n
\ee
Our technique typically provides the first $N=10$ terms of this expansion as we show next.
Last, we define the spectral function (or interacting local density of state) 
\be
A(\omega) = -\frac{1}{\pi}{\rm Im}[G^R(\omega)].
\ee
and the retarded self energy $\Sigma^R(\omega)$,
\begin{equation}
\label{self}
    G^R(\omega) = \frac{1}{\omega - \epsilon_d + i\Gamma - \Sigma^R(\omega)}
\end{equation}

\begin{table*}[t]
    \begin{ruledtabular}
    \begin{tabular}{||c|c|c|c|c|c|c|c|c||}
        $m$ & {$n=2$} (QMC) & $n=2$ (Yamada) & $n=2$ (Bethe) & {$n=4$} (QMC)  & $n=4$ (Yamada) & $n=4$ (Bethe) & {$n=6$} (QMC)  & $n=6$ (Bethe)\\
        \hline
        0 & $0 \pm 1\e{-5}$ & 0              &               & $0\pm2\e{-6}$  & 0              &               & $0\pm1\e{-7}$  & \\
        1 & 5.39(4)$\e{-2}$ & 5.3964$\e{-2}$ & 5.3964$\e{-2}$& 5.7(0)$\e{-4}$ & 5.6771$\e{-4}$ & 5.6482$\e{-4}$& 2.(1)$\e{-6}$  & 2.5119$\e{-6}$\\
        2 & 5.03(6)$\e{-2}$ & 5.0660$\e{-2}$ &               & 1.9(9)$\e{-3}$ & 2.0079$\e{-3}$ &               & 3.(1)$\e{-5}$  & \\
        3 & 3.67(5)$\e{-2}$ &                &               & 4.3(4)$\e{-3}$ &                &               & 1.(5)$\e{-4}$  & \\
        4 & 2.17(2)$\e{-2}$ &                &               & 6.2(4)$\e{-3}$ &                &               & 4.7(1)$\e{-4}$ & \\
    \end{tabular}
    \end{ruledtabular}
    \caption{
        \label{tab:se_coeffs}
        First coefficients $s_{n,m}$ (real) of the self-energy Taylor series
        $\Sigma(U,\omega)/\Gamma = \sum_{n,m} i^{m+1} s_{n,m}
        (U/\Gamma)^n(\omega/\Gamma)^m$ on the equilibrium symmetric model
        $\epsilon_d = 0$.  Coefficients in powers of $\omega$ have been
        obtained by fitting the bare data by a polynomial.  We find a good
        agreement with analytical calculations from
        Ref.~\onlinecite{Yamada1975} (2nd and 5th columns), as well as with
        Bethe ansatz exact calculations from Ref.~\onlinecite{Horvatic1985}
        (3rd, 6th and 8th columns).
    }
\end{table*}

%%%%%%%%%%%%%%%%%%%%%%%%%%%%%%%%%%%%%%%%%%%%%%%%%%%%%%%%%%%%%%%%%%%%%%%%%%%%%%%
\subsection{Numerical Results order by order}
We now present the numerical data obtained by sampling the kernel $K$. The left panels of Fig.~\ref{fig:kernels} shows an example of the bare data for the retarded
kernel $K^R(t)$ as they come out of the calculation for order $U^2$, $U^4$,
$U^6$ and $U^8$ (top to bottom). Note that the noise in these data is mostly
apparent, it corresponds to noise at very high frequency. This apparent noise
reflects the fact that we have binned the curve $K^R(t)$ into a very fine grid
($50 000$ grid points in this calculation). An even finer grid would show even
more apparent noise (since there would be even less Monte-Carlo points per grid
point). The corresponding cumulative function $\int_0^t K^R(u)du$ is however
noiseless as can be seen from  the example shown in
Fig.~\ref{fig:primitive_kernel} for $n=8$. 

The next step is to make a Fast Fourier Transform of $K^R_n(t)$ (not shown). The resulting $K^R_n(\omega)$ is relatively noisy at high frequency. Last, we obtain
$G^R_n(\omega)= g^R(\omega) K^R_n(\omega)$ for $n\geq 1$ (from Eq.~(\ref{eq:GA_from_KA})) as shown in the right panel of
Fig.~\ref{fig:kernels}. The factor $g^R(\omega)$, which decays at high
frequency, very efficiently suppresses the high frequency noise of the kernel
data. The same noise-reduction mechanism has been reported in \textit{e.g.} Ref.~\onlinecite{Gull2008}
in the context of auxiliary-field Monte-Carlo.
We emphasize one aspect of these data which is rather remarkable: even though the eight order contribution $G^R_8(\omega)$ is the result of an eight dimensional integral and is
more than four orders of magnitude smaller than the second order contribution
$G^R_2(\omega)$, it can be obtained with high precision (the error bars are of
the order of the thickness of the lines here). This is due to the recursive way
these integrals are calculated as discussed in Ref.~\onlinecite{Profumo2015}.

Using the definition Eq.~(\ref{self}) of the Self energy, we can obtain a recursive expression for $\Sigma^R_n$ in term of the Green's function expansion:
\begin{equation}
    \Sigma^R_n(\omega) = [g^R(\omega)]^{-2} G^R_n(\omega) - \sum_{k=1}^{n-1} \Sigma^R_{k}(\omega) G^R_{n-k}(\omega) g^R(\omega)^{-1}
\end{equation}
for $n>1$ with $\Sigma^R_1(\omega) = [g^R(\omega)]^{-2} G^R_1(\omega)$. The corresponding data is shown in Fig.~\ref{fig:SE_order_by_order} where we plot the coefficients
$\Sigma_n^R(\omega)$ for $n=2,4,6,8$ and $10$. The error bars increases with
the order $n$ which we attribute to the fact that, since the self energy only
contains one-particle irreducible diagrams, it is the subject of many
cancellations of terms. 
Indeed, one finds that the decay of $\Sigma^R_n(\omega)$ with $n$ is rather rapid with seven orders of magnitude between the first and the tenth order.

Our first benchmark uses a reference calculation made by Yamada\cite{Yamada1975}. The result at order $2$ is compared with the result of Yamada in the left panel 
of  Fig.~\ref{fig:SE_order_by_order} and found to be in excellent agreement. In
his seminal work Yamada also provided analytical calculations at order $2$ and
$4$ in the form of a low frequency expansion for the particle-hole symmetric
impurity,
\begin{equation}
    \Sigma^R(U,\omega) =\Gamma  \sum_{n,m} i^{m+1} s_{n,m} \left(\frac{\omega}{\Gamma}\right)^m  \left(\frac{U}{\Gamma}\right)^n
\end{equation}
Table~\ref{tab:se_coeffs} shows the results of Yamada ($m=1,2$ and $n=2,4$ ) as well as ours (obtained by fitting our numerical data at low frequency). We find
a good quantitative agreement with Yamada results. Yamada also provided numerical results at $n=4$ which are almost featureless and in very poor agreement
with our data.

\begin{figure}[h]
    \centering
    \includegraphics[width=7cm]{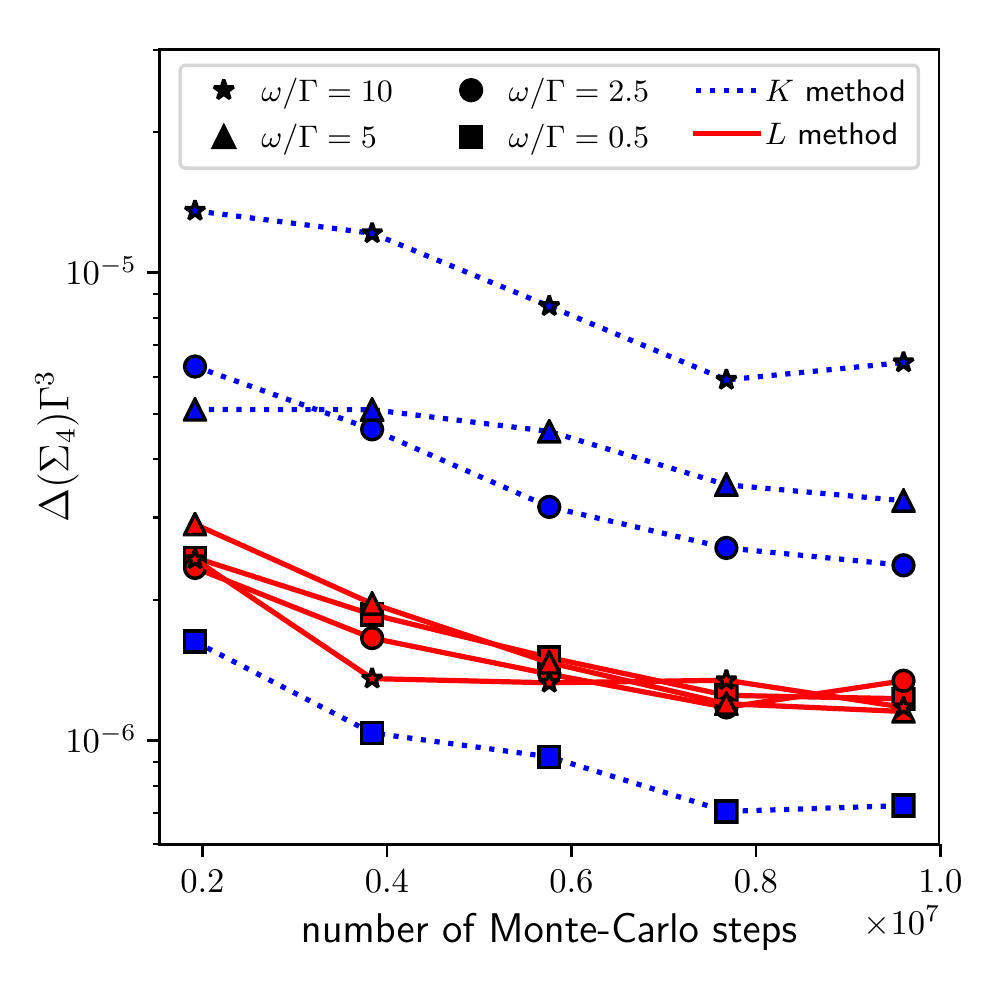}
    \caption{
        \label{fig:comparison_K_L}
        Statistical error of $\Sigma_4$ in the symmetric model at equilibrium,
        with increasing number of Monte-Carlo steps. The $K$ kernel method
        (dotted lines) and the $L$ kernel method (plain line) are compared for
        different frequencies (different symbols). The error with the $L$
        method is constant with frequencies, whereas the $K$ method accuracy
        worsen with increasing $\omega$. At large frequencies ($\omega >
        \Gamma$) the error is smaller when using the $L$ method.
    }
\end{figure}

Our second method uses the kernel $L$ in order to calculate the Green's function expansion. The bare data is very similar to the one obtained
with the kernel $K$ method. By construction, the reconstruction of $G$ with $L$ involves $G_n(\omega) \sim g(\omega)^2 L_{n-1}(\omega)$ so 
that the high frequency noise is expected to behave better with $L$ than with $K$ (the factor $ g(\omega)^2$ effectively suppresses the high frequency).
Fig.~\ref{fig:comparison_K_L} shows a comparison of the errors obtained on $\Sigma^R_4(\omega)$ using the two methods. We find that the error using the $L$ method is
essentially frequency independent while the error with the $K$ method depends
strongly on frequency. In most cases the $L$ method is preferred but at small
frequency, we have observed that the $K$ method can provide smaller error bars. 

%%%%%%%%%%%%%%%%%%%%%%%%%%%%%%%%%%%%%%%%%%%%%%%%%%%%%%%%%%%%%%%%%%%%%%%%%%%%%%%
\subsection{Numerical results for the spectral function}
\begin{figure}[h]
    \centering
    \includegraphics[width=8cm]{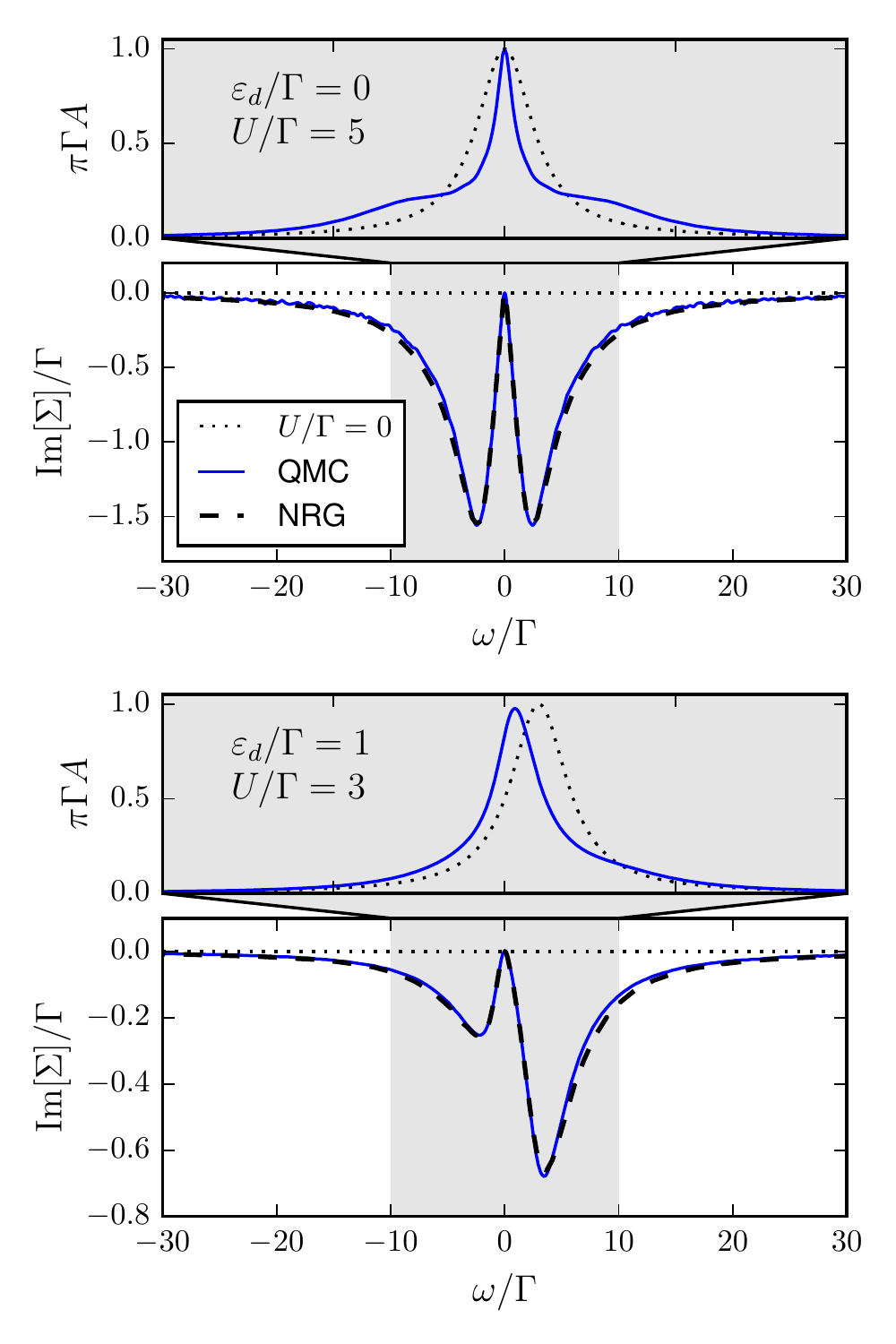}
    \caption{
    \label{fig:Bare_sum}
    Truncated series for the self energy $\Sigma^R(\omega)$ at $\epsilon_d= 0$ and $U= 5\Gamma$
    (second panel) and $\epsilon_d= \Gamma$ and $U= 3\Gamma$ (fourth panel) up to $N=10$ orders in perturbation theory.
    The first and third panels show the corresponding spectral function.
    The Monte-Carlo results (blue plain lines) are consistent with non-perturbative NRG
    calculations (dashed lines). The non-interacting situation is shown as
    dotted lines.
    Maximum integration time is $t_M = 20/\Gamma$.
    }
\end{figure}

Once the Green's function or self energy has been obtained up to a certain order, the last task is to extract the physics information from this expansion.
The most naive approach is to compute the truncated series up to a certain maximum order $N$,
\begin{equation}
\Sigma^R(U,\omega) \approx \sum_{n=1}^N \Sigma^R_n(\omega) U^n
\end{equation}
We find that the series has a convergence radius $U_c \approx 6\Gamma$ at the particle symmetry point $\epsilon_d=0$ while this
convergence radius decreases down to $U_c \approx 4\Gamma$ in the asymmetric case $\epsilon_d=\Gamma$. These convergence radii fix the maximum
strength of $U$ that one can study using the naive truncated series approach.

The data for the self energy (second and fourth panels) and corresponding spectral functions (first and third panels) are shown in  Fig.~\ref{fig:Bare_sum} for the symmetric
case (upper two panels, $\epsilon_d= 0$ and $U= 5\Gamma$) and asymmetric case (lower two panels, $\epsilon_d= \Gamma$ and $U= 3\Gamma$). For these values of interaction,
the error in our calculation is dominated by the finite truncation of the series (negligible error due to the statistical Monte-Carlo sampling) and is of the order of the line width.
Fig.~\ref{fig:Bare_sum} also shows the NRG results that we use to benchmark our
calculations and that are in excellent agreement with our data. The NRG
calculation is the same as in the companion paper\cite{Bertrand2019b}, where it
is described in details. Note that in order to obtain this agreement, the
precision of the NRG calculations had to be pushed much further than what is
typically done in the field indicating that the QMC method is very competitive,
in particular at large frequencies.

Qualitatively, the strength of interaction that could be reached using the
truncated series corresponds to the onset of the Kondo effect: one observes in
the upper panel of Fig.~\ref{fig:Bare_sum} that the Kondo peak starts to form
around $\omega = 0$, its width is significantly narrower than without
interaction and the premisses of the side peaks at $\pm U/2$ can be seen. In
order to observe well established Kondo physics, one must therefore go beyond
the convergence radius wall. This is in fact rather natural, the convergence
radius corresponds to poles or singularities in the complex $U$ plane which
themselves correspond to the characteristic energy scales of the system.
Getting past this ``convergence radius wall" is crucial and is the subject of
the companion article to the present manuscript.

%%%%%%%%%%%%%%%%%%%%%%%%%%%%%%%%%%%%%%%%%%%%%%%%%%%%%%%%%%%%%%%%%%%%%%%%%%%%%%%
\section{Conclusion}
\label{sec:conclusion}

In this article, we have presented a quantum Monte-Carlo algorithm that allows one to calculate the out of equilibrium Green's
functions of an interacting system, order by order in powers of the interaction coupling constant $U$.
We applied it to the Anderson model in the quantum dot geometry
and obtained up to $10$ orders of the Green's function and self-energy.
A detailed benchmark was also presented against NRG computations, after a simple summation of the series at weak coupling.
Our results were obtained at almost zero temperature, but we found that the method works equally well at finite
temperature or out-of-equilibrium.  It  works equally well for transient response to an interaction quench or at long time where a stationary regime
has been reached.

The method presented here has the great advantage to produce the perturbative expansion
{\it at infinite time}, {\it i.e. } directly in the steady state.
Its complexity is uniform in time: it does not grow at long time,
contrary to other QMC approaches.
The drawback, like any ``diagrammatic" QMC, is that we have just produced
the perturbative series of \textit{e.g.} the Green's function and the self-energy.
At weak coupling, we can simply sum it, as shown earlier in the benchmark.
However, at intermediate coupling, simply summing the series with partial sums
will fail. Most quantities have a finite radius $R$ of convergence in $U$:
for $U>R$, the series diverges.
In Ref.~\onlinecite{Profumo2015}, we showed how to use well-known conformal transformation resummation technique
to solve this problem and obtain density of particle on the dot vs $U$ up to $U=\infty$.
How to generalize this idea to make it work for real frequency Green's functions, and also to
control the amplification of stochastic noise due to such resummation
will be addressed in a separate publication \cite{Bertrand2019b}.

\section{Acknowledgement}

We thank Laura Messio for discussions at the early stage of this work.
We thank Serge Florens for interesting discussions and for providing NRG calculations.
The Flatiron Institute is a division of the Simons Foundation. 
We acknowledge financial support from the graphene Flagship (ANR FLagera GRANSPORT),
the French-US ANR PYRE and the French-Japan ANR QCONTROL.

\newpage

%%%%%%%%%%%%%%%%%%%%%%%%%%%%%%%%%%%%%%%%%%%%%%%%%%%%%%%%%%%%%%%%%%%%%%%%%%%%%%%
\appendix
%%%%%%%%%%%%%%%%%%%%%%%%%%%%%%%%%%%%%%%%%%%%%%%%%%%%%%%%%%%%%%%%%%%%%%%%%%%%%%%

\begin{widetext}

\section{Proof of the clusterization property of the kernel $K$}
\label{app:ClusterizationWProof}

In this appendix, we extend the proof of the clusterization property of Ref.~\onlinecite{Profumo2015}
for the Kernel $K$, $\bar K$ and $L$.
We want to show that, if some of the times $u_i$ are sent to infinity in the integral in Eq.~(\ref{eq:def_G_Keldysh_expansion}),
(\ref{eq:def_kernel_K}), (\ref{eq:def_Kernel_Kbar}) and~(\ref{eq:def_Kernel_KF}), 
the sum under the integral vanishes (while each determinant taken individually does not).
We will not try to prove here the stronger property that the integrals do indeed converge 
but we observe it empirically in the numerical computations.

Let us restart from the clusterization proof of Ref.~\onlinecite{Profumo2015} for
Eq.~(\ref{eq:def_G_Keldysh_expansion}) and examine the sum over the Keldysh indices:
\begin{equation}
S \equiv 
\sum_{\{a_k\}} \prod_{k=1}^n(-1)^{ a_k} 
\lists{(x,t,a), U_1, \ldots, U_{2n}}{(x',t',a'), U_1, \ldots, U_{2n}}
=
\sum_{\{a_k\}} \prod_{k=1}^n(-1)^{ a_k} 
\left|
\begin{matrix}
    {g}(X, X') &  {g}(X, U_1) & \hdots & {g}(X, U_{2n}) \\
    {g}(U_1, X') &  {g}(U_1, U_1) & \hdots & {g}(U_1, U_{2n}) \\
    \vdots &  \vdots & \ddots & \vdots \\
    {g}(U_{2n}, X') &  {g}(U_{2n}, U_1) &  \hdots & {g}(U_{2n}, U_{2n})
\end{matrix}
\right|
\end{equation}
If some $u_i$ are sent to infinity, we can relabel them $u_{p+1}, \dots, u_{n}$.
Since $g$ vanishes at large time (due to the presence of the bath), 
the determinants in the sum become diagonal by block
\begin{equation}
S \approx 
\sum_{\{a_k\}} \prod_{k=1}^n(-1)^{ a_k} 
\left|
\begin{matrix}
    g(X, X')       &  g(X, U_1)        & \hdots & g(X, U_{2p})       & 0                     & \hdots   & 0 \\
    g(U_1, X')     &  g(U_1, U_1)      & \hdots & g(U_1, U_{2p})     & 0                     & \hdots   & 0\\
    \vdots         & \vdots            & \ddots & \vdots             & \vdots                & \ddots   & \vdots \\
    g(U_{2p}, X')  &  g(U_{2p}, U_1)   & \hdots & g(U_{2p}, U_{2p})  & 0                     & \hdots   & 0 \\
    0              &  0                & \hdots & 0                  & g(U_{2p+1}, U_{2p+1}) & \hdots   & g(U_{2p+1}, U_{2n}) \\
    \vdots         & \vdots            & \ddots & \vdots             & \vdots                & \ddots   & \vdots \\
    0              &  0                & \hdots & 0                  & g(U_{2n}, U_{2p+1} )  & \hdots   & g(U_{2n}, U_{2n}) \\
\end{matrix}
\right|
\end{equation}
The upper-left determinant does not depend on $a_{p+1}, \dots, a_{n}$, so we can apply Eq.~(\ref{eq:ZisONE}) to the bottom-right determinant and 
the sum $S$ vanishes.

Let us now turn to the kernel $K$ defined in Eq.~(\ref{eq:def_kernel_K}).
The situation is slightly different.
First, with a simple relabelling, we can restrict ourselves to the case $p=1$ in Eq.~(\ref{eq:def_kernel_K}).
Let us first split the $U$ into two subsets. 
\begin{align}
S &=
\sum_{\{a_k\}} \prod_{k=1}^n(-1)^{ a_k} 
\left|
\begin{matrix}
    g(U_1, X')     &  g(U_1, U_2)      & \hdots & g(U_1, U_{2p})     & g(U_1, U_{2p+1})      & \hdots   & g(U_1, U_{2n})  \\
    \vdots         & \vdots            & \ddots & \vdots             & \vdots                & \ddots   & \vdots \\
    g(U_{2p}, X')  &  g(U_{2p}, U_2)   & \hdots & g(U_{2p}, U_{2p})  & g(U_{2p}, U_{2p+1})   & \hdots   & g(U_{2p}, U_{2n}) \\
    g(U_{2p+1}, X')&  g(U_{2p+1}, U_2) & \hdots & g(U_{2p+1}, U_{2p})& g(U_{2p+1}, U_{2p+1}) & \hdots   & g(U_{2p+1}, U_{2n}) \\
    \vdots         & \vdots            & \ddots & \vdots             & \vdots                & \ddots   & \vdots \\
    g(U_{2n}, X')  &  g(U_{2n}, U_2)   & \hdots & g(U_{2n}, U_{2p})  & g(U_{2n}, U_{2p+1} )  & \hdots   & g(U_{2n}, U_{2n}) \\
\end{matrix}
\right|
\end{align}
Some $u_i$ go to infinity. We distinguish two cases.
\begin{enumerate}
   \item If $u_1$ does not go to infinity, we can relabel the indices so that $u_{p+1}, \dots, u_{n}$ go to infinity. %$U_1, U_2$ and $X'$ stay at finite distance.
   \item
   If $u_1$ goes to infinity, we can relabel the indices so that $u_{1}, \dots, u_{p}$ go to infinity.
\end{enumerate}
In both cases, the upper-right part of the matrix vanishes and we get a block-trigonal determinant  
\begin{align*}
S & \approx 
   \sum_{\{a_k\}} \prod_{k=1}^n(-1)^{ a_k} 
\left|
\begin{matrix}
    g(U_1, X')     &  g(U_1, U_2)      & \hdots & g(U_1, U_{2p})     & 0                     & \hdots   & 0 \\ 
    \vdots         & \vdots            & \ddots & \vdots             & \vdots                & \ddots   & \vdots\\
    g(U_{2p}, X')  &  g(U_{2p}, U_2)   & \hdots & g(U_{2p}, U_{2p})  & 0                     & \hdots   & 0\\
    g(U_{2p+1}, X')&  g(U_{2p+1}, U_2) & \hdots & g(U_{2p+1}, U_{2p}) & g(U_{2p+1}, U_{2p+1}) & \hdots   & g(U_{2p+1}, U_{2n}) \\
    \vdots         & \vdots            & \ddots & \vdots             & \vdots                & \ddots   & \vdots \\
    g(U_{2n}, X')  &  g(U_{2n}, U_2)   & \hdots & g(U_{2n}, U_{2p})  & g(U_{2n}, U_{2p+1} )  & \hdots   & g(U_{2n}, U_{2n}) \\
\end{matrix}
\right|
\\
&= 
\left(
\sum_{a_1,\dots,a_p} \prod_{k=1}^p(-1)^{ a_k}
\lists{U_{1}, \ldots, U_{2p}}{X', U_2, \ldots, U_{2p}}
\right)
\times
\left(
\sum_{a_{p+1},\dots,a_n} \prod_{k=p+1}^n(-1)^{ a_k}
\lists{U_{2p+1}, \ldots, U_{2n}}{U_{2p+1}, \ldots, U_{2n}}
\right)
\end{align*}
since the first determinant does not depend on $a_{p+1}, \dots, a_{n}$.
The second term cancels 
because of (\ref{eq:ZisONE}).

\section{Expression of the kernel $L$ as a sum of Green's functions}
\label{app:L_as_gf}

We show here that the kernel $L$ can be expressed in terms of Green's
functions. Starting from its definition Eq.~(\ref{eq:def_Kernel_KF}), we follow
the same steps as in Section~\ref{sec:eq_of_motion}.
We first use the fact (due to determinant symmetry) that all terms of the sum
over $p$ have the same contribution:
\begin{multline}
L_{yx'z}^{ba'}(u,t') = (-1)^b
\sum_{n\geq 1}
\frac{i^n U^n}{n!} \int \prod_{k=1}^n du_k
\sum_{\{x_k, y_k\}}\sum_{\{a_k\}}
\left(
\prod_{k=1}^n (-1)^{a_k} V_{x_k y_k}(u_k)
\right)
\times \\
2n\,
\delta_c\Bigl((y,u,b), U_1\Bigr)
\lists{ (z,t',a'), U_1, U_2, \ldots, U_{2n}}{ (x',t',a'), (z,t',a'), U_2, \ldots, U_{2n}}
\end{multline}
Then we sum out the Dirac delta:
\begin{multline}
L_{yx'z}^{ba'}(u,t') = 2iU
\sum_{z'} V_{yz'}(u)
\sum_{n\geq 0}
\frac{i^n U^n}{n!}
\int \prod_{k=1}^n du_k
\sum_{\{x_k, y_k\}}\sum_{\{a_k\}}
\left(
\prod_{k=1}^n (-1)^{a_k} V_{x_k y_k}(u_k)
\right)
\times \\
\lists{ (z,t',a'), (y,u,b), (z',u,b), U_1, \ldots, U_{2n}}{ (x',t',a'), (z,t',a'), (z',u,b), U_1, \ldots, U_{2n}}
\end{multline}
The pattern of a 3-particle Green's function can be recognized:
\begin{equation}
L_{yx'z}^{ba'}(u,t') = 2iU
\sum_{z'} V_{yz'}(u)
E_{yx'zz'}^{ba'}(u, t')
\end{equation}
where $E$ is defined as:
\begin{equation}
    E_{yx'zz'}^{ba'}(u, t') \equiv (-i)^3 \left< T_{\rm c} 
    \hnd{c}(y,u,b)\hnd{c}^\dagger(x',t',a') 
    \left[ \hnd{c}^\dagger(z,t',a')\hnd{c}(z,t',a') - \alpha_z \right] 
    \left[\hnd{c}^\dagger(z',u,b)\hnd{c}(z',u,b) - \alpha_{z'}\right] 
    \right>
\end{equation}

\end{widetext}

%%%%%%%%%%%%%%%%%%%%%%%%%%%%%%%%%%%%%%%%%%%%%%%%%%%%%%%%%%%%%%%%%%%%%%%%%%%%%%%
\newpage
\bibliographystyle{apsrev}
\bibliography{../diag_MC_Anderson}{}

\end{document}